\documentclass{moriond}
\usepackage{bm}
\usepackage{amsmath}
\usepackage{multirow}

\def\Journal#1#2#3#4{{#1} {\bf #2}, #3 (#4)}

\def\PLB{{\em Phys. Lett.}  B}
\def\PRL{\em Phys. Rev. Lett.}
\def\PRD{{\em Phys. Rev.} D}

\def\be{\begin{equation}}
\def\ee{\end{equation}}
\def\bea{\begin{eqnarray}}
\def\eea{\end{eqnarray}}

\newcommand{\Photo}{}

\begin{document}
\vspace*{4cm}
\title{SEARCH FOR NEW PHYSICS SIGNALS VIA DOUBLY WEAK $\bm{B}$ DECAYS}

\author{FAISAL MUNIR BHUTTA$^{1,3}$, CAI DIAN L\"U$^{1,2,}$  }

\address{$^1$Institute of High Energy Physics, Chinese Academy of
Sciences, Beijing 100049, China\\
$^2$University of Chinese Academy of Sciences, Beijing 100049, China\\
$^3$Institute of Theoretical Physics, College of Applied Sciences,
Beijing University of Technology, Beijing 100124, China}

\maketitle\abstracts{
The doubly weak $b\to dd{\bar s}$ and $b\to ss{\bar d}$ processes are highly suppressed in the standard model that offer
the unique opportunity to explore new physics signals. The wrong sign decay $\smash{\overline B}^0\to K^+\pi^-$ mediated by the $b\to dd{\bar s}$  transition can be distinguished from the penguin decay $\smash{B}^0\to K^+\pi^-$, through time dependent measurement in experiments. 
We consider a model independent analysis
of $\smash{\overline B}^0\to K^+\pi^-$ decay, within the perturbative QCD approach and explore various effective
dimension-6 operators, in which large effects are possible. We also study the doubly weak exclusive process
in two example models namely Randall-Sundrum model with custodial protection and the bulk-Higgs Randall-Sundrum model.
A large and significant enhancement of the branching ratio, in comparison to the standard model, is observed after satisfying
all the relevant constraints on the parameter spaces of these models, which requires to be searched in future experiments.}

\section{Introduction}\label{sec1:intro}

Rare $B$ decays induced by flavor-changing neutral-current (FCNC) transitions come with an
interesting possibility of exploring the virtual effects from new physics (NP) beyond the
standard model (SM). Compared to radiative, leptonic and semi-leptonic rare $B$ decays, NP
investigations through purely hadronic $B$ decays pose added difficulty because of
relatively much larger theoretical hadronic uncertainties, which hinder to make any definite
conclusion for the presence of NP signals in the hadronic rare decays. Therefore we focus on
an alternate strategy proposed by Huitu {\it et al.} \cite{Huitu:1998vn}, which is
to search for the rare $b$ decay channels which have extremely small rates in the SM, so that
mere detection of such processes will be a clear signal of NP.

The doubly weak transitions, $b\to dd\bar s$ and $b\to ss\bar d$ are prototype processes which
occur via box diagrams and are highly suppressed in the SM, with branching ratios of approximately $\mathcal{O}(10^{-14})$
 and $\mathcal{O}(10^{-12})$, respectively. Both inclusive and
exclusive $B\to PP,PV, VV$ channels based on $b\to dd\bar s$ and $b\to ss\bar d$ transitions have been
investigated in several examples of NP models \cite{Fajfer:2006av,Pirjol:2009vz}. However, measuring these
two body doubly weak decays in experiments, is challenging, since in most cases they
mix with the ordinary weak decays through $B^0_{d,s}$-$\smash{\overline B}^0_{d,s}$ mixing or
$K^0$-$\smash{\overline K}^0$ mixing. In the case of $b \to dd\bar s$ transition, only three-body
decay $B^+\to \pi^+\pi^+K^-$ is searched experimentally and the LHCb collaboration has provided
the upper limit \cite{LHCb:2016rul} $\mathcal{B}(B^+\to \pi^+\pi^+K^-)<4.6\times 10^{-8}$.

We calculate the exclusive two body pure annihilation decay $\smash{\overline B}^0\to K^+\pi^-$ induced
by $b\to dd\bar s$  transition, within the perturbative QCD factorization approach (PQCD) and propose
to look for this wrong sign decay by performing a flavour-tagged time-dependent analysis of the right sign
decay $B^0\to K^+\pi^-$ with a large data sample, following reference \cite{Aubert:2004ei}.
For detail discussion, we refer the reader to \cite{Bhutta:2018tra}.

\section{$\smash{\overline B}^0 \to  K^+ \pi^- $ decay in the standard model}\label{sec2:SM}

\begin{figure}[ht]
\centerline{\includegraphics[width=11cm]{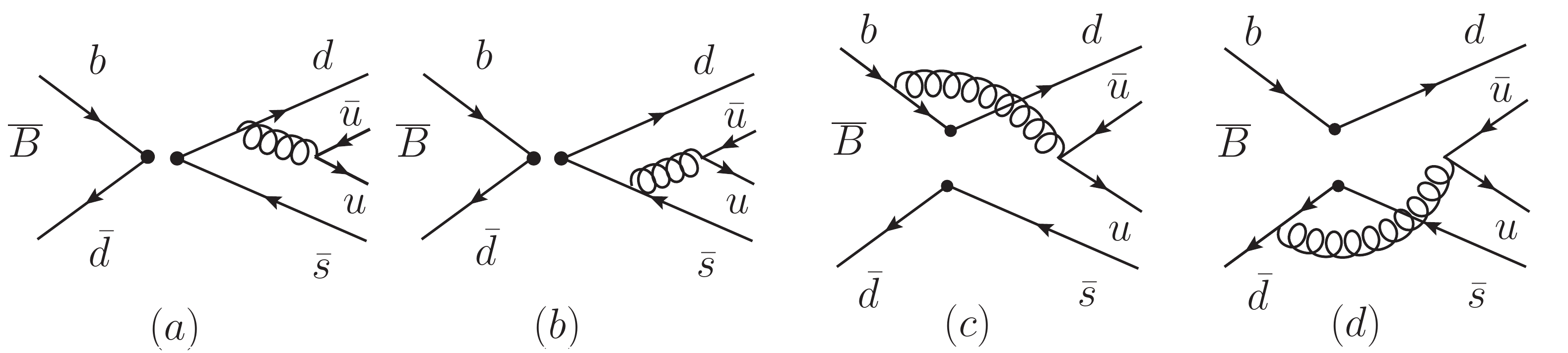}}
\caption{Leading order Feynman diagrams contributing to $\smash{\overline B}^0\to K^+\pi^-$ decay, ($a$) and
($b$) are the factorizable diagrams, ($c$) and ($d$) are the nonfactorizable diagrams.}
\label{fig:diagrams}
\end{figure}

In the SM, $b \to dd\bar s$ transition is both loop and CKM suppressed.
The local effective Hamiltonian for $b \to dd\bar s$ transition is given by:
\be
\mathcal{H}^{\text{SM}}=C^{\text{SM}}[(\bar d_L^{\alpha}\gamma^{\mu}b_L^{\alpha})
(\bar d_L^{\beta}\gamma_{\mu}s_L^{\beta})]
\label{eq:01}
\ee
with $C^{\text{SM}}$ involving the CKM matrix elements and the loops functions   given in \cite{Bhutta:2018tra}.
The exclusive $\smash{\overline B}^0\to K^+\pi^-$ decay, mediated by $b \to dd\bar s$ transition, proceeds
through the annihilation type Feynman diagrams which can be calculated in PQCD formalism\cite{Lu:2000em}.
Figure~\ref{fig:diagrams} shows four lowest order annihilation Feynman
diagrams for $\smash{\overline B}^0\to K^+\pi^-$ decay.
By solving these diagrams in PQCD factorization approach, we estimate the branching fraction in the SM:
\be
\mathcal{B}(\smash{\overline B}^0\to K^+\pi^-)^{\text{SM}}=1.0\times 10^{-19}.
\label{eq:02}
\ee

\section{Model independent analysis of the $\smash{\overline B}^0\to K^+\pi^-$ Decay}\label{sec3:NP}

To perform model independent analysis, we consider the most general local effective Hamiltonian with
all possible dimension-6 operators \cite{Grossman:1999av}:
\be
\mathcal{H}_{\text{eff}}^{\text{NP}}=\sum_{j=1}^5[C_j\mathcal{O}_j+\widetilde
{C}_j\widetilde{\mathcal{O}}_j],
\label{eq:GH}
\ee
where
\begin{equation}
\renewcommand{\arraystretch}{1.2}
\begin{array}{l@{\,\,\,\,\,\,\,\,}c@{\,\,\,\,\,\,\,\,}l}
\mathcal{O}_1=(\bar d_L\gamma_{\mu} b_L)(\bar d_L\gamma^{\mu} s_L), &
\mathcal{O}_2=(\bar d_R b_L)(\bar d_R s_L), & \mathcal{O}_3=(\bar d_R^{\alpha} b_L^{\beta})
(\bar d_R^{\beta}s_L^{\alpha}),\\
\mathcal{O}_4=(\bar d_R b_L)(\bar d_L s_R),     &
\mathcal{O}_5=(\bar d_R^{\alpha} b_L^{\beta})(\bar d_L^{\beta}s_R^{\alpha}). &
\end{array}
\label{eq:GO}
\end{equation}
The chirality flipped $\widetilde O_j$ operators can be written from $O_j$ by $L\leftrightarrow R$ exchange.
The NP beyond the SM can change the Wilson coefficient of operator $O_1$ and it can also provide non zero
Wilson coefficients for other new operators. These Wilson coefficients are not free parameters and are constrained
by $K^0-\smash{\overline K}^0$ and $B^0-\smash{\overline B}^0$ mixing parameters. In the case, where
NP only contributes to the local operator $\mathcal{O}_1$, similar to the SM, NP contributions are not allowed much
room due to a good agreement of the SM results for mixing observables with the experimental data. Among the remaining
nine non-standard operators, we consider each operator individually and calculate the corresponding decay width of the
$\smash{\overline B}^0\to K^+\pi^-$ decay given in case of $\mathcal{O}_{2-5}$ and $\widetilde{\mathcal{O}}_{1-5}$:
\begin{equation}
\renewcommand{\arraystretch}{1.2}
\begin{array}{l@{\,\,\,\,\,\,\,\,\,\,\,\,\,}l}
\Gamma_j=\frac{m_{B}^3}{64\pi}\left|F_{aj}\bigg[\frac{4}{3}C_j^{dd \bar s}\bigg]
+\mathcal{M}_{aj}\bigg[C_j^{dd\bar s}\bigg]\right|^2, &
\widetilde{\Gamma}_j=\frac{m_{B}^3}{64\pi}\left|F_{aj}\bigg[\frac{4}{3}\widetilde{C}_j^{dd \bar s}\bigg]
+\mathcal{M}_{aj}\bigg[-\widetilde{C}_j^{dd\bar s}\bigg]\right|^2,
\end{array}
\label{eq:amp}
\end{equation}
respectively. Index ``$j$'' corresponds to the operator number. The explicit expressions of $F_{aj}$ and ${\cal M}_{aj}$
are given in \cite{Bhutta:2018tra}. Further, we define the ratio $R$ between the branching fraction of the wrong sign decay
and the branching fraction of the right sign decay:
\be
R\equiv \frac{\mathcal{B}(\smash{\overline B}^0\to K^+\pi^-)}{\mathcal{B}(\smash{\overline B}^0\to K^-\pi^+)}.
\label{ratio}
\ee
Ratio $R$ can be directly measured in experiments. Assuming that the current experimental precision can probe $R$
to less than $10^{-3}$, we obtain the upper bound on the Wilson coefficient of each non-standard NP operator, presented
in Table \ref{tab:WC}.

\begin{table}[t]
\caption{Upper bounds on the Wilson coefficients of NP
operators for experimental precision $R<0.001$.}
\label{tab:WC}
\begin{center}
\begin{small}
\begin{tabular}{|c|c|c|c|}
\hline
Parameter   &  Allowed range (GeV$^{-2}$)  &  Parameter   &  Allowed range (GeV$^{-2}$)  \\
\hline
&   &  $\widetilde C_1$ & $<1.1\times 10^{-7}$ \\
$C_2$ & $<6.3\times 10^{-9}$  & $\widetilde C_2$ & $<6.8\times 10^{-9}$  \\
$C_3$ & $<5.1\times 10^{-8}$  & $\widetilde C_3$ & $<5.3\times 10^{-8}$ \\
$C_4$ & $<4.9\times 10^{-9}$ &  $\widetilde C_4$ & $<4.2\times 10^{-9}$\\
$C_5$ & $<1.6\times 10^{-6}$ & $\widetilde C_5$ & $<7.3\times 10^{-7}$\\
\hline
\end{tabular}
\end{small}
\end{center}
\end{table}

Next, we consider a NP scenario involving NP field $X$, with mass $M_X$, that carries a conserved quantum number.
In this NP example, it is possible to trivially satisfy mixing constraints through hierarchies among the NP couplings,
so that $b \to dd\bar s$ transition remains unbounded \cite{Pirjol:2009vz}. We consider four scenarios of NP
such that in S1 and S2, we suppose that NP matches onto the local operators $\mathcal{O}_1$
and $\mathcal{\widetilde{O}}_1$, respectively, while in S3 (S4), NP involves the linear
combination of local operators $\mathcal{O}_4$ ($\mathcal{O}_5$) and $\mathcal{\widetilde{O}}_4$ ($\mathcal{\widetilde{O}}_5$).
As $K^0-\smash{\overline K}^0$ and $B^0-\smash{\overline B}^0$ mixing bounds
do not constrain $M_{X}$ in this case, we assume two cases of NP scale with $M_{X}= 1$ TeV and $M_{X}= 10$ TeV, respectively,
and obtain the resulting PQCD prediction for the ratio $R_X$ in each scenario. The results for both cases along with $R_{\text{SM}}$
are listed in Table \ref{tab:b}.

\begin{table}[htb]
\caption{Ratio $R_X$ in case of NP carrying conserved charge along with $R_{\text{SM}}$.}\label{tab:b}
\begin{center}
\begin{small}
\begin{tabular}{ | c | c | c | c | c | c | }
\hline
\multirow{2}{*}{Scenarios}   &  \multicolumn{4}{c|}{$R_X$} & \multirow{2}{*}{$R_{\text{SM}}$} \\ \cline{2-5}
& $M_X$ (TeV) & Case-I  & $M_X$ (TeV) & Case-II  &\\ \hline
S1 &\multirow{4}{*}{1} &0.085  & \multirow{4}{*}{10}  & $8.5\times 10^{-6}$ & $6.8\times 10^{-15}$\\ 
S2 &  & 0.074 &     &$7.3\times 10^{-6}$ & \\ 
S3 & &  55    &     & $0.005$ & \\ 
S4 & &  0.002 &     & $1.9\times 10^{-7}$ & \\
\hline
\end{tabular}
\end{small}
\end{center}
\end{table}

\section{$ \smash{\overline B}^0 \to K^+ \pi^- $ in Randall-Sundrum Models}\label{sec:4RScmodel}

$\smash{\overline B}^0\to K^+\pi^-$ decay in the custodial Randall-Sundrum $(\text{RS}_c)$ model
\cite{Randall:1999ee,Blanke:2008zb} receives tree level contributions from the lightest Kaluza-Klein (KK) gluons
$\mathcal{G}^{(1)}$, KK photon $A^{(1)}$ and new heavy electroweak (EW) gauge bosons $(Z_H, Z^{\prime})$. $Z$ contributions
protected through discrete $P_{LR}$ symmetry and $\Delta F=2$ contributions from Higgs boson exchanges are negligible.
In the bulk-Higgs Randall-Sundrum (RS) model, $\smash{\overline B}^0\to K^+\pi^-$ decay results from the tree-level
exchanges of KK gluons, KK photons, $Z$ and the Higgs boson including their KK excitations and from the extended scalar
fields $\phi^{Z(n)}$. Starting with the local effective Hamiltonian of Eq.~(\ref{eq:GH}) and assigning a common name
$M_{g^{(1)}}$ to the masses of the lightest KK gauge bosons, we calculate the Wilson coefficients in both the RS models at
$\mu=\mathcal{O}(M_{g^{(1)}})$. For explicit expressions we refer to \cite{Bhutta:2018tra}. The dominant contribution,
in both RS models, comes from the KK gluons, while in $\text{RS}_c$ model $Z_H, Z^{\prime}$ bosons try to compete with KK
gluons contributions.

\begin{figure}
\begin{minipage}{0.33\linewidth}
\centering
\includegraphics[width=0.95\linewidth]{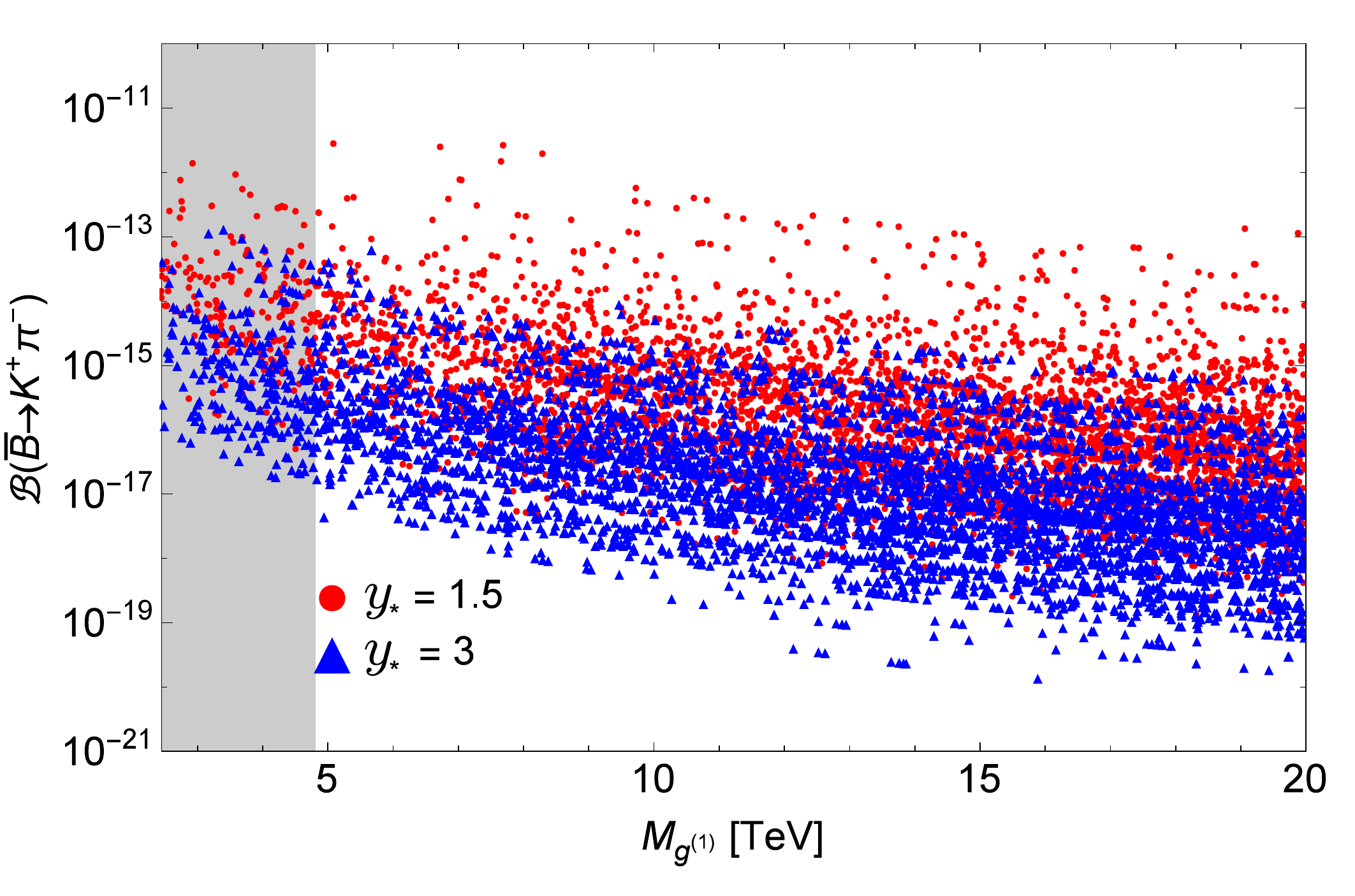}\\
{\scriptsize (a) $\text{RS}_c$ model,}
\end{minipage}
\hfill
\begin{minipage}{0.32\linewidth}
\centering
\includegraphics[width=0.95\linewidth]{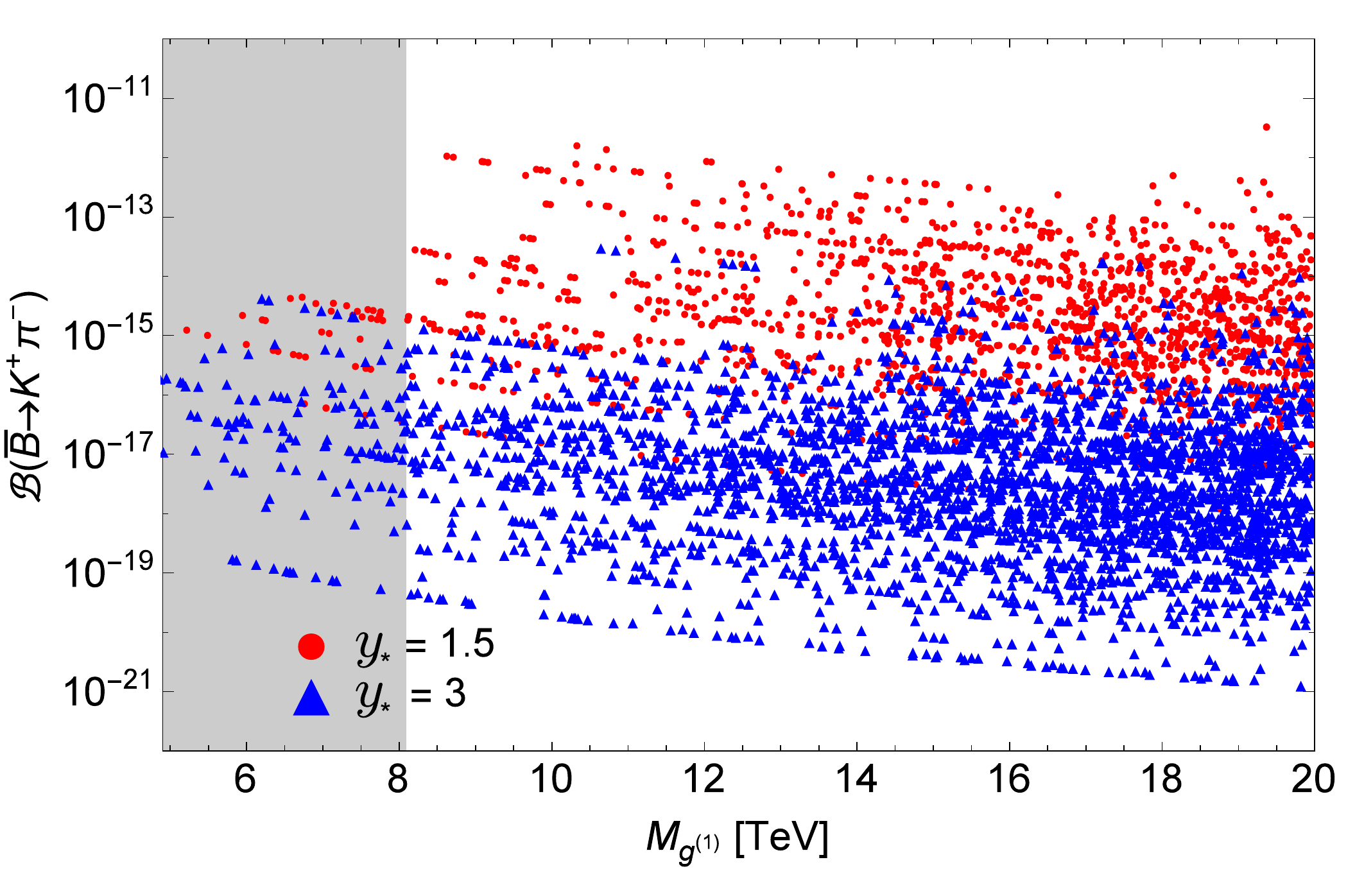}\\
{\scriptsize (b) $\beta=1$ bulk-Higgs,}
\end{minipage}
\hfill
\begin{minipage}{0.32\linewidth}
\centering
\includegraphics[width=0.95\linewidth]{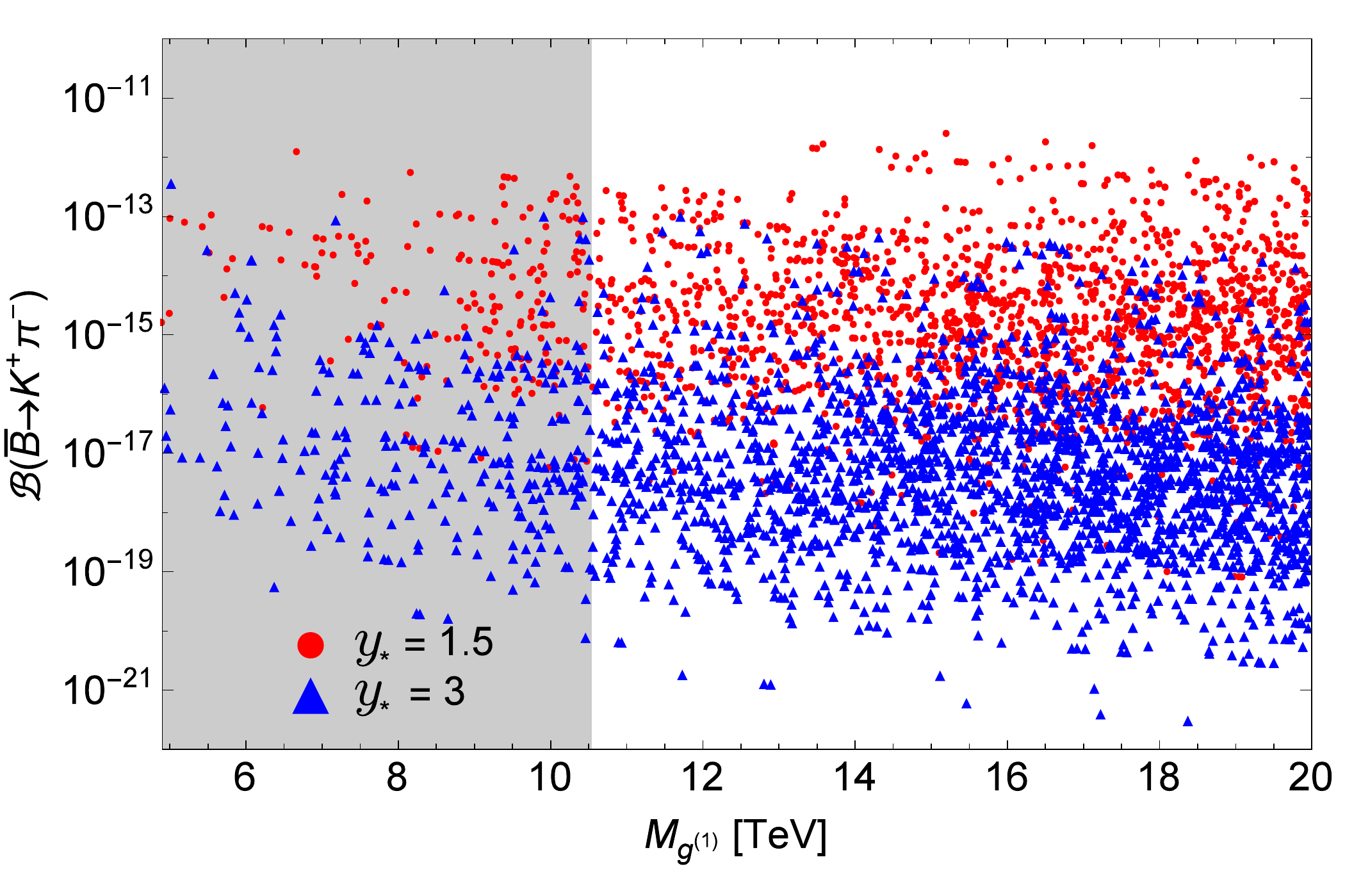}\\
{\scriptsize (c) $\beta=10$ bulk-Higgs.}
\end{minipage}
\caption{$\smash{\overline B}^0\to K^+\pi^-$ branching fraction vs. the KK gluon mass $M_{g^{(1)}}$
with two different values of $y_{\star}$, (a) in the $\text{RS}_c$ model and (b), (c) in the bulk-Higgs
RS model with $\beta=1$ and $\beta = 10$. The gray regions are excluded by the analysis of electroweak
precision experiments.}
\label{fig:bH}
\end{figure}

Two sets of data points are generated, by using the strategy employed in
\cite{Blanke:2008zb,Lu:2016pfs,Nasrullah:2018vky}, which belong to anarchic 5D Yukawa
coupling matrices with $y_{\star}=1.5$ and $3$. $y_{\star}$ defines the maximum allowed value for the elements of the
5D Yukawa matrices. In the bulk-Higgs RS model, we use two different values of parameter $\beta$, which belong to different
localization of the Higgs field along the extra dimension. Figure~\ref{fig:bH} shows a range of PQCD predictions for
the branching ratio of the $\smash{\overline B}^0\to K^+\pi^-$ decay as a function of $M_{g^{(1)}}$ with two different
values of $y_{\star}$, in both the RS models, after simultaneously incorporating the $\Delta m_K$, $\epsilon_K$ and $\Delta m_{B_d}$
constraints. $y_{\star}=1.5$ and $y_{\star}=3$ cases are shown as red and blue scatter points, respectively. The gray
shaded areas are excluded by the EW precision data. It is evident from Figure~\ref{fig:bH}(a) that in the $\text {RS}_c$ model,
a maximum increase of six orders of magnitude in the branching ratio, compared to SM result, is achievable for $y_{\star}=1.5$ case.
While from Figure~\ref{fig:bH}(b) and \ref{fig:bH}(c), in the bulk-Higgs RS model, $\smash{\overline B}^0\to K^+\pi^-$ decay
can get a maximum enhancement of five and six orders of magnitude for $y_{\star}=3$ and $y_{\star}=1.5$ value, respectively
with both values of $\beta=1$ and $\beta = 10$.

\section{Conclusions}\label{sec5:con}

Doubly weak $\smash{\overline B}^0\to K^+\pi^-$ decay, mediated by $b \to dd\bar s$ transition,
is studied in detail within the PQCD framework and a method is proposed to distinguish it from the right sign
decay $B^0\to K^+\pi^-$, by the time-dependent measurement of neutral B decays in $B^0-\smash{\overline B}^0$ mixing.
In the model independent analysis, constraints on the Wilson coefficients of the new physics dimension-6 operators are derived
for a specific experimental precision of the ratio $R$, while very large predictions for the ratio $R_X$
in different NP scenarios involving NP with a conserved charge are obtained due to the hierarchies among the NP couplings.
In two variants of the RS model, after satisfying all the relevant constraints, a maximum enhancement of five to six orders
of magnitude, in the decay rate of $\smash{\overline B}^0\to K^+\pi^-$, is possible for different parameter values, which
leaves this decay free for the search of NP in future experiments.

\section*{Acknowledgments}

We thank Ying Li and Yue-Hong Xie for collaboration on this topic. The work is partly supported by National Science Foundation of China (11521505 and 11621131001).

\end{document}